\documentclass[%
prc
,twocolumn%
,superscriptaddress
,amsmath,amssymb,aps,nobibnotes,
nofootinbib
,preprintnumbers
]{revtex4-1}

\usepackage[dvipdfmx]{graphicx}
\usepackage{bm}
\usepackage{braket}
\usepackage{dcolumn}
\usepackage{color}

\begin{document}
\title{Functional renormalization-group calculation 
of the equation of state of one-dimensional nuclear matter
 inspired by the Hohenberg--Kohn theorem}

\author{Takeru Yokota}
\email{tyokota@ruby.scphys.kyoto-u.ac.jp}
\author{Kenichi Yoshida}
\email{kyoshida@ruby.scphys.kyoto-u.ac.jp}
\affiliation{Department of Physics, Faculty of Science, Kyoto University, Kyoto 606-8502, Japan}
\author{Teiji Kunihiro}
\email{kunihiro@yukawa.kyoto-u.ac.jp}
\affiliation{Yukawa Institute for Theoretical Physics, 
Kyoto University, Kyoto 606-8502, Japan}


\begin{abstract}
We present the first successful functional renormalization 
group(FRG)-aided density-functional (DFT) calculation of the 
equation of state (EOS) of an infinite nuclear matter (NM) 
in (1+1)-dimensions composed of spinless nucleons.
We give a formulation to describe infinite matters
in which the 'flowing' chemical potential is introduced to control
the particle number during the flow.
The resultant saturation energy
of the NM coincides with that obtained by the Monte-Carlo method
within a few percent. 
Our result demonstrates that the FRG-aided DFT 
can be as powerful as any other 
methods in quantum many-body theory. 
\end{abstract}

\maketitle

\section{Introduction\label{sec:intro}}

The Hohenberg-Kohn (HK) theorem
\cite{hoh64}
tells us that the problems of quantum many-body systems
can be formulated solely in terms of 
the particle density $\rho(x)$ 
without the many-particle wave function.
A formalism based on this theorem
is the density functional theory (DFT)
utilizing the energy density functional (EDF) $E[\rho]$.
The DFT is used in various fields
including quantum chemistry,  
condensed matter physics, and nuclear physics:
Thanks to the practical methods
based on the Kohn-Sham formalism
\cite{koh65}, 
DFT has become a powerful method 
to analyze the properties of ground states;
see Refs.~\cite{koh99,kry14,zan15,jon15} for an overview.
Methods to investigate excited states
such as the time-dependent density functional theory \cite{run84}
have also been developed \cite{ull12,nak16}.

It should be, however, noted that the HK theorem
only guarantees the existence of the EDF $E[\rho]$ 
which could be minimized to obtain the exact ground-state density and energy,
but does not provide 
any theoretical prescription to construct $E[\rho]$ itself.
Thus most practical  calculations
utilize $E[\rho]$ that is constructed 
in a semi-empirical way.
Hence developing a systematic method to derive
 $E[\rho]$ from microscopic Hamiltonian still remains as a fundamental
problem in the field of quantum many-body theory.

A clue of this fundamental problem may be provided 
by the notion of effective field theory developed in 
quantum field theory. Indeed the two-particle point-irreducible
(2PPI) effective action \cite{ver92}
can lead to an energy density functional written in terms of 
$\rho$ for which the HK theorem naturally emerges \cite{pol02,sch04}.
A nice point with the effective-action approach is that
an established powerful computational machinery is now available, which is called
 the functional renormalization group (FRG)
method \cite{wet93,weg73,wil74,pol84}.
In this method, 
the quantum fluctuations
are gradually taken into account from an ultraviolet 
to infrared scale by solving the one-parameter
flow equation of the scale-dependent effective action, and hence 
a coarse-grained effective action is eventually obtained;
see Refs.~\cite{ber02,gie12,paw07} for reviews.
Since the 2PPI effective action is 
a generalization of the energy density functional,
the FRG method formulated for the 2PPI action
possibly gives a formal foundation to DFT and provides
a long-desired method for constructing the density functional from 
a microscopic Hamiltonian,
as initiated by Polonyi, Sailer and Schwenk
\cite{pol02,sch04}.
Such an observation has lead to a notion of DFT-RG or 
2PPI-FRG method \cite{bra12,kem13,kem17a,kem17b,lia18},
which is a quite attractive scheme for solving the 
fundamental problem in the field of the DFT: In a pioneering work \cite{kem13},
accurate ground state energies of simple toy models in quantum mechanics 
 are obtained within the fourth-order truncation:
A recent paper \cite{lia18} proposed an efficient method
  to incorporate the higher-order correlations, and the method was  applied to a $0$-dimensional
quartic  model successfully.
The subsequent analysis of  
 a one-dimensional system composed of a finite number of particles
motivated by the nuclear saturation problem
\cite{kem17a,kem17b}, however,
showed that the second-order truncation 
only gives 
a 30\% accuracy in comparison with the result of the Monte-Carlo simulation \cite{ale89}.
Possible improvement of the result may be obtained by an incorporation of 
the higher-order correlation functions as 
suggested in the demonstration in a 0-dimensional model \cite{lia18}.

In this paper, we apply the DFT-RG scheme to an infinite uniform system. 
Our point is that an infinite system with a definite particle density may be well described 
with first few correlation functions while  rarefied systems
are interaction-dominating systems and higher correlation functions may play  significant roles.
Needless to say, an infinite uniform system is a fundamentally important system for understanding
 many-body physics 
and indeed the local density approximation for $E[\rho]$
was found to be unexpectedly successful~\cite{koh65}.
We give a DFT-RG formalism for infinite matters
in which we introduce a 'flowing' chemical potential
to control the flow of the particle number caused by switching on 
the inter-particle interaction.
Then we calculate the equation of state (EOS) of an infinite nuclear matter (NM) 
in (1+1)-dimensions composed of spinless nucleons 
as in Refs.~\cite{ale89,kem17a}.
Starting from the two-'nucleon' interaction constructed in Ref.~\cite{ale89}
where the saturation curve of the one-dimensional NM is obtained by 
the Monte-Carlo simulation,
we solve the flow equation for the 2PPI effective action with some reasonable 
truncation.
We show that the resultant density functional $E[\rho]$ 
nicely gives the saturation energy, i.e. the minimum of the energy
derived by the EOS with respect to the density,
that coincides with that of the Monte Carlo method \cite{ale89}.

This paper is organized as follows: 
In Sec.~\ref{sec:form}, our formalism is shown.
We introduce the flowing chemical potential
to control the particle number during the flow
and derive the DFT-RG flow equation
for infinite uniform systems 
with a definite particle number.
In Sec.~\ref{sec:demo},
we apply our formalism to a (1+1)-dimensional spinless
nuclear matter. The results of the density dependence of the
ground state energy, i.e. the equation of state
is shown in this section.
Section~\ref{sec:conc} is devoted to the conclusion.

\section{Formalism\label{sec:form}}
In this section, we show our formalism 
to analyze ground state energies
of one-dimensional continuum matters 
composed of spinless fermions
in the framework of DFT-RG.

We consider (1+1)-dimensional spinless fermions with a two-body interaction $U$.
We employ the imaginary-time finite-temperature formalism for convenience.
Then the action in the units such that mass of a fermion is $1$ reads
\begin{align}
	S[\psi^*,\psi]&=
	\int_{\chi}
	\psi^* (\chi_\epsilon)\left(
	\partial_\tau-\frac{\partial^2_x}{2}
	\right)\psi(\chi)
	\notag
	\\
	&
	+\frac{1}{2}\int_{\chi,\chi'}
	U_{\rm 2b}(\chi,\chi')
	\psi^{*}(\chi_\epsilon)\psi^{*}(\chi'_\epsilon)
	\psi(\chi')\psi(\chi),
	\label{eq:acti}
\end{align}
where $\chi:=(\tau, x)$, 
$\chi_\epsilon:=(\tau+\epsilon, x)$ with a positive infinitesimal $\epsilon$,
$\int_{\chi}:=\int_\tau\int_x
:=\int_{0}^{\beta}d\tau \int dx$ with an inverse temperature $\beta$,
and $U_{\rm 2b}(\chi,\chi'):=U_{\rm 2b}(\chi-\chi'):=\delta(\tau-\tau')U(x-x')$.
The imaginary times of the fermion fields $\psi$ and $\psi^*$ at the same point
are infinitesimally different, which comes from the construction
of the path integral formalism \cite{alt10}.

Following the prescription in Ref.~\cite{pol02,sch04},
we introduce the regulated interaction 
$U_{\rm 2b, \lambda}(\chi)$ ($U_{\lambda}$(x))
such that $U_{\rm 2b, \lambda=0}(\chi)=0$
and $U_{\rm 2b, \lambda=1}(\chi)=U_{\rm 2b}(\chi)$
($U_{\lambda=0}(x)=0$ and $U_{\lambda=1}(x)=U(x)$).
Then the regulated action is defined
in terms of $U_{\rm 2b, \lambda}(\chi)$:
\begin{align}
	S_\lambda [\psi^*,\psi]&=
	\int_{\chi}
	\psi^* (\chi)\left(
	\partial_\tau-\frac{\partial^2_x}{2}
	\right)\psi(\chi)
	\notag
	\\
	&
	+\frac{1}{2}\int_{\chi,\chi'}
	U_{\rm 2b, \lambda}(\chi,\chi')
	\psi^{*}(\chi_\epsilon)\psi^{*}(\chi'_\epsilon)
	\psi(\chi')\psi(\chi).
	\label{eq:actr}
\end{align}
This action becomes that for free particles at $\lambda=0$ and Eq.~\eqref{eq:acti} at $\lambda=1$.
The parameter $\lambda \in [0,1]$ is 
interpreted as the flow parameter from 
the free to interacting system.
In this paper, we choose 
$U_{\rm 2b,\lambda}(\chi,\chi')=\lambda U_{\rm 2b}(\chi,\chi')$.

The EDF $E_\lambda[\rho]$ realizing
the HK theorem can be defined
in terms of the 2PPI effective action~\cite{fuk94},
which is defined as follows:
\begin{align}
	\Gamma_{\lambda}[\rho]=\sup_{J}
	\left( \int_\chi J(\chi)\rho(\chi)
	-W_{\lambda}[J]\right),
	\label{eq:effe}
\end{align}
where $\rho(\chi)$ is the density field,
$W_\lambda[J]:=\ln Z_\lambda[J]$
is the generating functional for the connected density
correlation functions and 
$Z_{\lambda}[J]
=\int \mathcal{D}\psi^* \mathcal{D}\psi 
\exp(-S_{\lambda}[\psi^*,\psi]
+\int_{\chi}J(\chi)\rho_\psi(\chi))$
is the generating functional
for the correlation functions of the density field
$\rho_\psi(\chi)=\psi^*(\chi_\epsilon)\psi(\chi)$.
To see the correspondence of the 2PPI effective action
to the EDF, we consider the variational problem
of $\Gamma_\lambda[\rho]$ with a fixed number of particles.
The stationary condition of the effective action
determines the behavior of the expectation value
of the density field.
Under the constraint that the number of particles is fixed,
we should minimize 
$I_\lambda[\rho]:=\Gamma_\lambda[\rho]-\mu_\lambda\int_\chi\rho(\chi)$
with respect to $\rho(\chi)$.
In general, the Fermi energy, and the particle number, depend on the interaction
and change during 
the flow\footnote{We are grateful to
Jean-Paul Blaizot for pointing this out.}. 
We thus have introduced a $\lambda$-dependent
Lagrange multiplier, or physically 
a 'flowing' chemical potential,
$\mu_\lambda$ to control the particle number during the flow.
The ground state density $\rho_{\rm gs,\lambda}(\chi)$
satisfies the stationary condition:
\begin{align}
	\Gamma^{(1)}_\lambda[\rho_{\rm gs,\lambda}](\chi)
	=J_{\rm sup}[\rho_{\rm gs,\lambda}](\chi)
	=\mu_\lambda,
	\label{eq:stat}
\end{align}
where $J_{\rm sup}[\rho](\chi)$ is $J(\chi)$ maximizing the right hand side of Eq.~\eqref{eq:effe} and
\begin{align*}
\Gamma_\lambda^{(n)}[\rho](\chi_1,\cdots,\chi_n)
:=\frac{\delta^{n} \Gamma[\rho]}{\delta \rho(\chi_1)\cdots\delta \rho(\chi_n)}.
\end{align*}
Because $-W_\lambda[\mu_\lambda]/\beta$ is the grand potential, we have
$\Gamma_\lambda[\rho_{\rm gs,\lambda}]=
\mu_\lambda\int_\chi \rho_{\rm gs,\lambda}(\chi) - W_{\lambda}[\mu_\lambda]
= \beta F_{\lambda}$,
where $F_\lambda$ is the Helmholtz free energy.
At the zero temperature limit $\beta\to \infty$,
$\Gamma_\lambda[\rho_{\rm gs,\lambda}]/\beta$
becomes the ground state energy $E_{\rm gs,\lambda}$
because $F_\lambda$ can be written as
$F_\lambda=-\beta^{-1}\ln \sum_n \exp(-\beta E_{n,\lambda})$,
where $\lbrace E_{n,\lambda} \rbrace$ is the energy eigenvalues
of the system
and satisfies $E_{\rm gs,\lambda}=E_{0,\lambda}<E_{1,\lambda}<\cdots$.
Therefore $\Gamma_\lambda[\rho]$ can be related 
to the EDF:
\begin{align}
	E_\lambda[\rho]=\lim_{\beta\to\infty} \frac{\Gamma_\lambda[\rho]}{\beta}.
	\label{eq:edfd}
\end{align}
$\rho_{\rm gs,\lambda}(\chi)$
is uniquely determined because of the convexity
of the effective action
except for the case when
the eigenvalues of
$\Gamma^{(2)}_\lambda[\rho_{\rm gs,\lambda}](\chi,\chi')$
do not satisfy positivity,
which implies that the ground state is unstable \cite{pol02}.

The flow equation of $\Gamma_\lambda[\rho]$ is derived 
by differentiating Eq.~\eqref{eq:effe}
with respect to $\lambda$:
$
\partial_\lambda \Gamma_\lambda[\rho]
=-(\partial_\lambda W_\lambda)[J_{\rm sup,\lambda}[\rho]].
$
The right-hand side of this flow equation becomes
\begin{align}
	-&(\partial_\lambda W_\lambda)[J_{\rm sup,\lambda}[\rho]]
	\notag
	\\
	=&
	\frac{1}{2}\int_{\chi,\chi'}
	\dot{U}_{\rm 2b,\lambda}(\chi,\chi')
	\braket{
	\psi^* (\chi_\epsilon)
	\psi^* (\chi'_\epsilon)
	\psi(\chi')
	\psi(\chi)
	}_\rho
	\notag
	\\
	=&
	\frac{1}{2}\int_{\chi,\chi'}
	\dot{U}_{\rm 2b,\lambda}(\chi,\chi')
	\left(
	\braket{\rho_\psi(\chi)}_\rho
	\braket{\rho_\psi(\chi')}_\rho
	\right.
	\notag
	\\
	&+
	\left.
	W_{\lambda}^{(2)}[J_{\rm sup,\lambda}[\rho]](\chi,\chi')
	-
	\delta(x-x')
	\braket{\rho_\psi(\chi)}_\rho
	\right),
	\label{eq:wflo}
\end{align}
where $\dot{U}_{\rm 2b, \lambda}(\chi,\chi'):=\partial_\lambda U_{\rm 2b, \lambda}(\chi,\chi')$,
\begin{align*}
	\braket{\cdots}_\rho
	:=&
	\int \mathcal{D}\psi^* \mathcal{D}\psi
	\cdots
	\frac{{\rm e}^{
	-S_{\lambda}[\psi^*,\psi]
	+\int_{\chi}J_{\rm sup,\lambda}[\rho](\chi)\rho_\psi(\chi)}}{Z_{\lambda}[J_{\rm sup,\lambda}[\rho]]},
\end{align*}
and
\begin{align*}
	W_\lambda^{(n)}[J](\chi_1,\cdots,\chi_n)
	:=
	\frac{\delta^{n} W_\lambda [J]}
	{\delta J(\chi_1)\cdots\delta J(\chi_n)}.
\end{align*}
To derive Eq.~\eqref{eq:wflo}, we have used 
$U_{\rm 2b}(\chi,\chi') \sim \delta(\tau-\tau')$
and the canonical commutation relation:
$\langle
\psi^* (\tau+\epsilon,x)
(\psi^* (\tau+\epsilon,x')\psi(\tau,x')+\psi(\tau+\epsilon,x')\psi^* (\tau,x'))
\psi(\tau,x)
\rangle_\rho=\delta(x-x')\langle\rho_\psi(\chi)\rangle_\rho$.
We note that $\braket{\cdots}_\rho$ gives averages for imaginary-time-ordered operator products
and that the density--density correlation function 
$W_\lambda^{(2)}[J_{\rm sup,\lambda}[\rho]](\chi,\chi')$ at $\tau=\tau'$ should be interpreted as
$
\lim_{\tau\to\tau'}\lim_{\epsilon\to +0}
(\braket{
\psi^* (\chi_\epsilon)
\psi(\chi)
\psi^* (\chi'_\epsilon)
\psi(\chi')}_\rho
-
\braket{\psi^* (\chi_\epsilon)\psi(\chi)}_\rho
\braket{\psi^* (\chi'_\epsilon)\psi(\chi')}_\rho
)
$
where the limit $\tau\to \tau'$ is taken after
the limit $\epsilon\to +0$, i.e. $|\tau'-\tau|>\epsilon$.
By use of the relations
$\braket{\rho_\psi(\chi)}_\rho=W^{(1)}_{\lambda}[J_{\rm sup,\lambda}](\chi)=\rho(\chi)$
and
\begin{align*}
	&\int_{\chi'}
	\Gamma^{(2)}_\lambda[\rho](\chi,\chi')
	W^{(2)}_\lambda[J_{\rm sup,\lambda}[\rho]](\chi',\chi'')
	\\
	&=\int_{\chi'}\frac{\delta J_{\rm sup,\lambda}[\rho](\chi)}{\delta \rho(\chi')}
	\frac{\delta \rho(\chi')}{\delta J_{\rm sup,\lambda}[\rho](\chi'')}=\delta(\chi,\chi'')
\end{align*}
the flow equation can be written in term of $\Gamma_\lambda[\rho]$ \cite{sch04, kem17a}:
\begin{align}
	\partial_\lambda \Gamma_\lambda [\rho]
	=&
	\frac{1}{2}\int_{\chi,\chi'}
	\dot{U}_{\rm 2b,\lambda}(\chi,\chi')
	\left(
	\rho (\chi)
	\rho(\chi')
	\right.
	\notag
	\\
	&+
	\left.
	\Gamma_{\lambda}^{(2)-1}[\rho](\chi, \chi')
	-\rho(\chi)\delta(x-x')
	\right).
	\label{eq:flow}
\end{align}
where $\Gamma_\lambda^{(2)-1}[\rho](\chi,\chi')$
is the inverse of $\Gamma_\lambda^{(2)}[\rho](\chi,\chi')$.


In principle, the functional flow equation~\eqref{eq:flow}
with the effective action $\Gamma_{0}[\rho]$
for the free fermions
gives the effective action $\Gamma_{1}[\rho]$ 
for the interacting fermions.
In general, however, some approximation is needed 
for the practical use of Eq.~\eqref{eq:flow}.
Here, we employ the vertex expansion:
\begin{align*}
	\Gamma_{\lambda}[\rho]=&
	\Gamma_\lambda[\rho_{\rm gs,\lambda}]
	+\sum_{n=1}^{\infty}
	\int_{\chi_{1}}\cdots\int_{\chi_{n}}
	\Gamma_\lambda^{(n)}[\rho_{\rm gs,\lambda}](\chi_1,\cdots,\chi_n)
	\\
	&\times
	(\rho(\chi_1)-\rho_{\rm gs,\lambda}(\chi_1))\cdots
	(\rho(\chi_n)-\rho_{\rm gs,\lambda}(\chi_n)).
\end{align*}
Up to the second order expansion,
Eq.~\eqref{eq:flow} is rewritten as the following
flow equations:
\begin{widetext}
\begin{align}
	\partial_\lambda \Gamma_\lambda[\rho_{\rm gs, \lambda}]
	=&
	\int_\chi
	\Gamma_\lambda^{(1)}[\rho_{\rm gs, \lambda}](\chi)
	\partial_\lambda \rho_{\rm gs, \lambda}
	+
	\frac{1}{2}\int_{\chi,\chi'}
	\dot{U}_{\rm 2b,\lambda}(\chi,\chi')
	\left(
	\rho_{\rm gs, \lambda}(\chi)
	\rho_{\rm gs, \lambda}(\chi')
	\right.
	\notag
	\\
	&+
	\left.
	\Gamma_{\lambda}^{(2)-1}[\rho_{\rm gs,\lambda}](\chi, \chi')
	-\rho_{\rm gs, \lambda}(\chi)\delta(x-x')
	\right),
	\label{eq:0thf}
	\\
	\partial_\lambda \Gamma_\lambda^{(1)}[\rho_{\rm gs, \lambda}]
	(\chi)
	=&
	\int_{\chi'}
	\Gamma^{(2)}_\lambda[\rho_{\rm gs, \lambda}](\chi,\chi')
	\partial_\lambda \rho_{\rm gs, \lambda}(\chi')
	+
	\int_{\chi'}
	\dot{U}_{\rm 2b,\lambda}(\chi,\chi')
	\rho_{\rm gs, \lambda}(\chi')
	-
	\frac{1}{2}
	\dot{U}_{\lambda}(0)
	\notag
	\\
	&-
	\frac{1}{2}\int_{\chi_1,\chi_2,\chi_3,\chi_4}
	\dot{U}_{\rm 2b,\lambda}(\chi_4,\chi_1)
	\Gamma_{\lambda}^{(2)-1}[\rho_{\rm gs,\lambda}](\chi_1, \chi_2)
	\Gamma_{\lambda}^{(3)}[\rho_{\rm gs,\lambda}](\chi_2, \chi_3,\chi)
	\Gamma_{\lambda}^{(2)-1}[\rho_{\rm gs,\lambda}](\chi_3, \chi_4),
	\label{eq:1stf}
	\\
	\partial_\lambda \Gamma_\lambda^{(2)}[\rho_{\rm gs, \lambda}]
	(\chi,\chi')
	=&
	\int_{\chi_1}
	\Gamma^{(3)}_\lambda[\rho_{\rm gs, \lambda}](\chi,\chi',\chi_1)
	\partial_\lambda \rho_{\rm gs, \lambda}(\chi_1)
	+
	\dot{U}_{\rm 2b,\lambda}(\chi,\chi')
	\notag
	\\
	&-
	\frac{1}{2}\int_{\chi_1,\cdots,\chi_4}
	\dot{U}_{\rm 2b,\lambda}(\chi_4,\chi_1)
	\Gamma_{\lambda}^{(2)-1}[\rho_{\rm gs,\lambda}](\chi_1, \chi_2)
	\Gamma_{\lambda}^{(4)}[\rho_{\rm gs,\lambda}](\chi_2, \chi_3,\chi,\chi')
	\Gamma_{\lambda}^{(2)-1}[\rho_{\rm gs,\lambda}](\chi_3, \chi_4)
	\notag
	\\
	&
	+\int_{\chi_1,\cdots,\chi_6}
	\dot{U}_{\rm 2b,\lambda}(\chi_4,\chi_1)
	\Gamma_{\lambda}^{(2)-1}[\rho_{\rm gs,\lambda}](\chi_1, \chi_2)
	\Gamma_{\lambda}^{(3)}[\rho_{\rm gs,\lambda}](\chi_2, \chi_3,\chi)
	\notag
	\\
	&
	\times
	\Gamma_{\lambda}^{(2)-1}[\rho_{\rm gs,\lambda}](\chi_3, \chi_4)
	\Gamma_{\lambda}^{(3)}[\rho_{\rm gs,\lambda}](\chi_4, \chi_5,\chi')
	\Gamma_{\lambda}^{(2)-1}[\rho_{\rm gs,\lambda}](\chi_5, \chi_6).
	\label{eq:2ndf}
\end{align}
These flow equation can be simplified by rewriting in terms of 
the connected correlation functions:
\begin{align*}
	G^{(n)}_\lambda(\chi_1,\cdots,\chi_n)
	=
	W_\lambda^{(n)}
	[J_{\rm sup}[\rho_{\rm gs,\lambda}]](\chi_1,\cdots,\chi_n).
\end{align*}
$\Gamma^{(n)}_\lambda[\rho_{\rm gs,\lambda}]$ 
is related to the connected correlation functions 
with the following relation:
\begin{align*}
	\Gamma_\lambda^{(n)}[\rho_{\rm gs,\lambda}](\chi_1,\cdots,\chi_n)
	=
	\left.
	\left(
	\prod_{i=1}^{n-2}
	\int_{\chi'_i}
	W_\lambda^{(2)-1}[J](\chi_i,\chi'_i)\frac{\delta}{\delta J(\chi'_i)}
	\right)
	W_\lambda^{(2)-1}[J](\chi_{n-1}, \chi_n)
	\right|_{J=J_{\rm sup}[\rho_{\rm gs,\lambda}]},
\end{align*}
which is derived from the following identity:
\begin{align}
	\frac{\delta}{\delta \rho (\chi)}
	=
	\int_{\chi'}
	\frac{\delta J_{\rm sup,\lambda}[\rho](\chi')}{\delta \rho(\chi)}
	\frac{\delta}{\delta J_{\rm sup,\lambda}(\chi')}
	=
	\int_{\chi'}	
	W_{\lambda}^{(2)-1}[J_{\sup,\lambda}[\rho]](\chi,\chi')
	\frac{\delta}{\delta J_{\rm sup,\lambda}(\chi')}.
\end{align}
The relations between 
$\Gamma_{\lambda}^{(2,3,4)}[\rho_{\rm gs, \lambda}]$ 
and the connected correlation functions read
\begin{align*}
	\Gamma_{\lambda}^{(2)}[\rho_{\rm gs, \lambda}](\chi_1,\chi_2)
	=&
	G^{(2)-1}_{\lambda}(\chi_1,\chi_2)
	\\
	\Gamma_{\lambda}^{(3)}[\rho_{\rm gs, \lambda}](\chi_1,\chi_2,\chi_3)
	=&
	-\int_{\chi'_1,\chi'_2,\chi'_3}
	G^{(3)}_{\lambda}(\chi'_1,\chi'_2,\chi'_3)
	G^{(2)-1}_{\lambda}(\chi_1,\chi'_1)
	G^{(2)-1}_{\lambda}(\chi_2,\chi'_2)
	G^{(2)-1}_{\lambda}(\chi_3,\chi'_3)
	\\
	\Gamma_{\lambda}^{(4)}[\rho_{\rm gs, \lambda}](\chi_1,\chi_2,\chi_3.\chi_4)
	=&
	-\int_{\chi'_1,\chi'_2,\chi'_3.\chi'_4}
	G^{(4)}_{\lambda}(\chi'_1,\chi'_2,\chi'_3,\chi'_4)
	G^{(2)-1}_{\lambda}(\chi_1\chi'_1)
	G^{(2)-1}_{\lambda}(\chi_2,\chi'_2)
	G^{(2)-1}_{\lambda}(\chi_3,\chi'_3)
	G^{(2)-1}_{\lambda}(\chi_4,\chi'_4)
	\\
	&+\int_{\chi'_1,\chi'_2,\chi'_3}
	G^{(2)-1}_{\lambda}(\chi_2,\chi'_2)
	G^{(2)-1}_{\lambda}(\chi_3,\chi'_3)
	G^{(3)}_{\lambda}(\chi'_1,\chi'_2,\chi'_3)
	\\
	&\times G^{(2)-1}_{\lambda}(\chi'_5,\chi'_1)
	G^{(3)}_{\lambda}(\chi'_4,\chi'_5,\chi'_6)
	G^{(2)-1}_{\lambda}(\chi_1,\chi'_4)
	G^{(2)-1}_{\lambda}(\chi_4,\chi'_6)
	\\
	&+\int_{\chi'_1,\chi'_2,\chi'_3}
	G^{(2)-1}_{\lambda}(\chi_4,\chi'_2)
	G^{(2)-1}_{\lambda}(\chi_3,\chi'_3)
	G^{(3)}_{\lambda}(\chi'_1,\chi'_2,\chi'_3)
	\\
	&\times G^{(2)-1}_{\lambda}(\chi'_5,\chi'_1)
	G^{(3)}_{\lambda}(\chi'_4,\chi'_5,\chi'_6)
	G^{(2)-1}_{\lambda}(\chi_1,\chi'_4)
	G^{(2)-1}_{\lambda}(\chi_2,\chi'_6)
	\\
	&+\int_{\chi'_1,\chi'_2,\chi'_3}
	G^{(2)-1}_{\lambda}(\chi_2,\chi'_2)
	G^{(2)-1}_{\lambda}(\chi_4,\chi'_3)
	G^{(3)}_{\lambda}(\chi'_1,\chi'_2,\chi'_3)
	\\
	&\times G^{(2)-1}_{\lambda}(\chi'_5,\chi'_1)
	G^{(3)}_{\lambda}(\chi'_4,\chi'_5,\chi'_6)
	G^{(2)-1}_{\lambda}(\chi_1,\chi'_4)
	G^{(2)-1}_{\lambda}(\chi_3,\chi'_6).
\end{align*}
By use of these relations,
Eqs.~\eqref{eq:0thf}-\eqref{eq:2ndf} are rewritten
as follows:
\begin{align}
	\partial_\lambda \Gamma_{\lambda}[\rho_{\rm gs,\lambda}]
	=&
	\int_\chi
	\mu_\lambda
	\partial_\lambda\rho_{\rm gs,\lambda}(\chi)
	+
	\frac{1}{2}
	\int_{\chi,\chi'}
	\dot{U}_{\rm 2b,\lambda}(\chi,\chi')
	\left(
	\rho_{\rm gs, \lambda}(\chi)
	\rho_{\rm gs, \lambda}(\chi')
	+
	G_{\lambda}^{(2)}(\chi,\chi')
	-\rho_{\rm gs,\lambda}(\chi')\delta(x'-x)
	\right),
	\label{eq:fflo}
	\\
	\partial_\lambda \rho_{\rm gs,\lambda}(\chi)
	=&
	-\frac{1}{2}\int_{\chi_1,\chi_2}
	\dot{U}_{\rm 2b,\lambda}(\chi_1,\chi_2)
	G_{\lambda}^{(3)}(\chi_2,\chi_1,\chi)
	\notag
	\\
	&+
	\int_{\chi_1}G_{\lambda}^{(2)}(\chi,\chi_1)
	\left(
	\partial_\lambda \mu_\lambda
	-
	\int_{\chi_2}
	\dot{U}_{\rm 2b,\lambda}(\chi_1,\chi_2)
	\rho_{\rm gs,\lambda}(\chi_2)
	+
	\frac{1}{2}\dot{U}_\lambda(0)
	\right),
	\label{eq:rflo}
	\\
	\partial_\lambda G_{\lambda}^{(2)}(\chi,\chi')
	=&
	-\int_{\chi_1,\chi_2}
	\dot{U}_{\rm 2b,\lambda}(\chi_1,\chi_2)
	\left(
	G_{\lambda}^{(2)}(\chi,\chi_1)
	G_{\lambda}^{(2)}(\chi_2,\chi')
	+\frac{1}{2}
	G_{\lambda}^{(4)}(\chi_2,\chi_1,\chi,\chi')
	\right)
	\notag
	\\
	&+
	\int_{\chi_1}G_{\lambda}^{(3)}(\chi,\chi',\chi_1)
	\left(
	\partial_\lambda \mu_\lambda
	-
	\int_{\chi_2}
	\dot{U}_{\rm 2b,\lambda}(\chi_1,\chi_2)
	\rho_{\rm gs,\lambda}(\chi_2)
	+
	\frac{1}{2}\dot{U}_\lambda(0)
	\right).
	\label{eq:g2fl}
\end{align}
\end{widetext}
Here, we have used
Eqs.~\eqref{eq:stat} and  \eqref{eq:edfd}.
These flow equations determine the behavior of the free energy $\Gamma_\lambda[\rho_{\rm gs,\lambda}]=\beta F_\lambda$,
the ground state density $\rho_{\rm gs,\lambda}(\chi)$ and the density--density correlation function $G_\lambda^{(2)}(\chi,\chi')$
under a given $\lambda$-dependent chemical potential $\mu_\lambda$.

For the case of free fermion $\lambda=0$,
the ground state of the system is homogeneous.
In this paper, we assume that the homogeneity of the system
remains even if $\lambda\neq 0$, i.e. the transition to
an inhomogeneous state does not occur even if the interaction
is switched on.
In general, the switching on of the interaction at $\lambda\neq 0$ changes the density
as represented in Eq.~\eqref{eq:rflo}.
However, we can compensate this effect of the interaction
by choosing an appropriate $\mu_\lambda$
and realize $\partial_\lambda \rho_{\rm gs,\lambda}=0$ 
in the case of the homogeneous system\footnote{Our idea can be 
extended to the case of inhomogeneous systems
by use of the $x$-dependent chemical potential $\mu_\lambda(x)$.
However, it would be impossible to fix $\rho_{\rm gs,\lambda}(\chi)$ to
an arbitral density such as those not satisfying 
the v-representability \cite{lev82,lie83,eng83}.}.
We employ the momentum representation for convenience
to discuss how to choose $\mu_\lambda$.
In the momentum representation,
Eqs.~\eqref{eq:fflo}-\eqref{eq:g2fl} in the case of homogeneous states read
\begin{widetext}
\begin{align}
	\partial_\lambda \frac{\Gamma_{\lambda}[\rho_{\rm gs,\lambda}]}{\beta V}
	=&
	\mu_\lambda
	\partial_\lambda\rho_{\rm gs,\lambda}
	+
	\frac{1}{2}\tilde{U}(0)\rho_{\rm gs,\lambda}^2
	+
	\frac{1}{2}
	\int_{p}
	\tilde{U}(p)
	\left(
	T\sum_{\omega}
	\tilde{G}^{(2)}_{\lambda}(P)
	-
	\rho_{\rm gs,\lambda}
	\right),
	\label{eq:fflp}
	\\
	\partial_\lambda \rho_{\rm gs,\lambda}
	=&
	-\frac{1}{2}\int_{P}
	\tilde{U}(p)
	\tilde{G}_{\lambda}^{(3)}(P,-P)
	+
	\tilde{G}_{\lambda}^{(2)}(0)
	\left(
	\partial_\lambda \mu_\lambda
	-
	\tilde{U}(0)
	\rho_{\rm gs,\lambda}
	+
	\frac{1}{2}U(0)
	\right),
	\label{eq:rflp}
	\\
	\partial_\lambda \tilde{G}_{\lambda}^{(2)}(P)
	=&
	-
	\tilde{U}(p)
	G_{\lambda}^{(2)}(P)^2
	-\frac{1}{2}
	\int_{P'}\tilde{U}(p')\tilde{G}^{(4)}_{\lambda}(P',-P',P)
	+
	G_{\lambda}^{(3)}(P,-P)
	\left(
	\partial_\lambda \mu_\lambda
	-
	\tilde{U}(0)
	\rho_{\rm gs,\lambda}
	+
	\frac{1}{2}U(0)
	\right).
	\label{eq:g2fp}
\end{align}
\end{widetext}
Here we have used $U_{\rm 2b,\lambda}(\chi,\chi')=\lambda U(x-x')\delta(\tau-\tau')$ and
introduced the volume of the system $V$ and the Fourier transformations
$\tilde{U}(p):=\int_{x}U(x)e^{-ipx}$
and
$(2\pi)^2\delta(P_1+\cdots+P_n)
\tilde{G}_\lambda^{(n)}(P_1,\cdots,P_{n-1})
:=
\int_{\chi_1,\cdots,\chi_n}
e^{-i(P_1\cdot\chi_1
+\cdots+P_n\cdot\chi_n)}
G_\lambda^{(n)}(\chi_1,\cdots,\chi_{n})$,
where $P_{i}:=(\omega_{i},p_i)$ is a vector of
a Matsubara frequency and a momentum.
We have introduced the short hands
$\int_{p}:=\int dp/(2\pi)$ 
and $\int_{P}:=\int_p T\sum_{\omega}$.
Then $\partial_\lambda \rho_{\rm gs,\lambda}=0$ is realized if the flow of $\mu_\lambda$ is set as follows:
\begin{align}
	\partial_\lambda \mu_\lambda
	=&
	\tilde{U}(0)
	\rho_{\rm gs,\lambda}
	-
	\frac{U(0)}{2}
	+
	\int_{P}
	\frac{\tilde{U}(p)\tilde{G}_{\lambda}^{(3)}(P,-P)}{2\tilde{G}_{\lambda}^{(2)}(0)}.
	\label{eq:mufl}
\end{align}
We note that $\tilde{G}_\lambda^{(2)}(0)$ should be
interpreted as the $p$ limit of $\tilde{G}_\lambda^{(2)}(P)$:
$\tilde{G}_\lambda^{(2)}(0)=\lim_{p\to 0}\tilde{G}^{(2)}_\lambda(0,p)$,
because the Matsubara frequency $\omega$ is discrete.
The $p$ limit of the $\tilde{G}^{(2)}_\lambda(P)$ is the static 
particle-density susceptibility and generally nonzero,
while $\lim_{p\to 0}\tilde{G}^{(2)}_\lambda(P)=0$ with a finite frequency
\cite{for75,kun91,fuj04}.
This is in contrast to the case of a finite number of particles in
a finite box \cite{kem17a}, where density correlation functions with vanishing
frequency and momentum were interpreted as the $\omega$ limit, i.e.,
not only $\int dx_{i} G_\lambda^{(n)}(\chi_1,\cdots,\chi_n)=0$
but also $\int_{\chi_{i}} G_\lambda^{(n)}(\chi_1,\cdots,\chi_n)=0$
with $i\in \lbrace 1,\cdots,n\rbrace$ were used to derive the flow equations.


In this paper, 
we focus on the zero temperature limit.
In the zero temperature limit with the condition Eq.~\eqref{eq:mufl},
Eqs.~\eqref{eq:fflp} and \eqref{eq:g2fp} becomes as follows:
\begin{widetext}
\begin{align}
	\partial_\lambda \overline{E}_{\rm gs,\lambda}
	=&
	\frac{\rho_{\rm gs,0}}{2} \tilde{U}(0)
	+
	\frac{1}{2\rho_{\rm gs,0}}
	\int_{p}
	\tilde{U}(p)
	\left(
	\int_{\omega_{\rm}}
	\tilde{G}^{(2)}_{\lambda}(P)
	-
	\rho_{\rm gs,0}
	\right),
	\label{eq:eflow}
	\\
	\partial_\lambda \tilde{G}_{\lambda}^{(2)}(P)
	=&
	-\tilde{U}(p) \tilde{G}_{\lambda}^{(2)}(P)^2
	-
	\frac{1}{2}
	\int_{P'}
	\tilde{U}(p')
	\tilde{G}^{(4)}_{\lambda}(P',-P',P)
	+
	\int_{P'}
	\frac{
	\tilde{U}(p')
	\tilde{G}_{\lambda}^{(3)}(P',-P')
	\tilde{G}_{\lambda}^{(3)}(P,-P)
	}{2\tilde{G}_{\lambda}^{(2)}(0)}
	\label{eq:gtfl},
\end{align}
\end{widetext}
where we have introduced the energy per particle 
$\overline{E}_{\rm gs,\lambda}=\lim_{\beta\to 0}\Gamma_{\lambda}[\rho_{\rm gs,\lambda}]/(\beta \int_x \rho_{\rm gs,0})$ and the shorthand 
$\int_\omega=\int d\omega/(2\pi)$.


The flow equation for $G^{(n)}_\lambda$
generally depends on $G^{(m\leq n+2 )}_\lambda$, 
which means that an infinite series of coupled flow equations emerges.
We avoid to treat such an infinite series of coupled flow equations
by ignoring the flows of $G_\lambda^{(3\leq n)}$.
However, we do not simply substitute 
$\tilde{G}_\lambda^{(3,4)}$ for $\tilde{G}_{0}^{(3,4)}$ in Eq.~\eqref{eq:gtfl}.
Such a simple replacement breaks a constraint 
for multi-particle distribution functions 
imposed by the Pauli blocking:
By use of the canonical commutation relation,
the $n$-particle distribution function
$f_{n,\lambda}(x_1,\cdots,x_n)=\lim_{\epsilon\rightarrow +0}\braket{\psi^*(\epsilon,x_1)\cdots\psi^*(\epsilon,x_n)\psi(0,x_n)\cdots\psi(0,x_1)}_{\rho_{\rm gs,\lambda}}$
is related to the connected correlation functions
$G_\lambda^{(m\leq n)}$.
Because of the Pauli blocking, the distribution function
satisfies $f_{n,\lambda}(x_1,\cdots,x_n)=0$ 
if $x_i=x_j$ for $i\neq j$.
In the case of $n=2$, the relation between $f_{2,\lambda}(x_1,x_2)$
and $G_\lambda^{(2)}$ is derived 
in the same same manner as the aforementioned derivation
of Eq.~\eqref{eq:wflo}:
\begin{align*}
	&f_{2,\lambda}(x_1,x_2)
	\notag
	\\
	=&\braket{\rho_{\psi}(0,x_1)\rho_{\psi}(0,x_2)}_{\rho_{\rm gs,\lambda}}
	-\delta(x_1-x_2)
	\braket{\rho_{\psi}(0,x_1)}_{\rho_{\rm gs,\lambda}}
	\notag
	\\
	=&
	G_\lambda^{(2)}(0,x_1,0,x_2)
	+\rho_{\rm gs,\lambda}(x_1)
	\rho_{\rm gs,\lambda}(x_2)
	\notag
	\\
	&-\delta(x_1-x_2)\rho_{\rm gs,\lambda}(x_1).
\end{align*}
Then the condition imposed by the Pauli blocking reads
\begin{align*}
	&\lim_{x_1\rightarrow x_2} G_\lambda^{(2)}(0,x_1,0,x_2)
	\notag
	\\
	=&
	\lim_{x_1\rightarrow x_2}
	\left(\delta(x_1-x_2)\rho_{\rm gs,\lambda}(x_1)
	-\rho_{\rm gs,\lambda}(x_1)\rho_{\rm gs,\lambda}(x_2)
	\right).
\end{align*}
In our case, the right-hand side of this condition
does not depend on $\lambda$:
$\lim_{x_1\rightarrow x_2} \partial_\lambda G_\lambda^{(2)}(0,x_1,0,x_2)=0$,
because $\partial_\lambda \rho_{\rm gs,\lambda}=0$.
Therefore, from Eq.~\eqref{eq:gtfl}, 
the following condition should be satisfied:
\begin{align}
	&-
	\int_{P}\tilde{U}(p) \tilde{G}_{\lambda}^{(2)}(P)^2
	-
	\frac{1}{2}
	\int_{P,P'}
	\tilde{U}(p')
	\tilde{G}^{(4)}_{\lambda}(P',-P',P)
	\notag
	\\
	&+
	\int_{P,P'}
	\frac{
	\tilde{U}(p')
	\tilde{G}_{\lambda}^{(3)}(P',-P')
	\tilde{G}_{\lambda}^{(3)}(P,-P)}
	{2\tilde{G}_{\lambda}^{(2)}(0)}
	=0
	\label{eq:paul}
\end{align}
This condition, however, is broken by 
the simple substitution 
$\tilde{G}_\lambda^{(3,4)}$ for $\tilde{G}_{0}^{(3,4)}$.
To respect the condition Eq.~\eqref{eq:paul},
we approximate the second and third terms in the right-hand side
of Eq.~\eqref{eq:gtfl} as follows \cite{kem17a}:
\begin{align}
	&-
	\frac{1}{2}
	\int_{P'}
	\tilde{U}(p')
	\tilde{G}^{(4)}_{\lambda}(P',-P',P)
	\notag
	\\
	&+
	\int_{P'}
	\frac{
	\tilde{U}(p')
	\tilde{G}_{\lambda}^{(3)}(P',-P')
	\tilde{G}_{\lambda}^{(3)}(P,-P)}
	{2\tilde{G}_{\lambda}^{(2)}(0)}
	\notag
	\\
	\approx
	&
	c_\lambda\left(-
	\frac{1}{2}
	\int_{P'}
	\tilde{U}(p')
	\tilde{G}^{(4)}_{0}(P',-P',P)
	\right.
	\notag
	\\
	&
	+
	\left.
	\int_{P'}
	\frac{
	\tilde{U}(p')
	\tilde{G}_{0}^{(3)}(P',-P')
	\tilde{G}_{0}^{(3)}(P,-P)
	}{2\tilde{G}_{0}^{(2)}(0)}
	\right)
	\label{eq:appr}
\end{align}
where $c_\lambda$ is the factor to preserve the condition Eq.~\eqref{eq:paul}:
\begin{align*}
	c_\lambda
	&=
	\int_{P}\tilde{U}(p) \tilde{G}_{\lambda}^{(2)}(P)^2
	\\
	&\times\left(
	-
	\frac{1}{2}
	\int_{P',P''}
	\tilde{U}(p')
	\tilde{G}^{(4)}_{0}(P',-P',P'')
	\right.
	\\
	&
	+
	\left.
	\int_{P',P''}
	\frac{
	\tilde{U}(p')
	\tilde{G}_{0}^{(3)}(P',-P')
	\tilde{G}_{0}^{(3)}(P'',-P'')}
	{2\tilde{G}_{0}^{(2)}(0)}
	\right)^{-1}.
\end{align*}
At $\lambda=0$, we have $c_0=1$.
In our approximation, 
the contribution of diagrams such as multi-pair creations
is not included. Such diagrams are important to
investigate the spectral properties of one-dimensional fermion
systems \cite{pus06,teb07,ima12},
which is beyond the scope of this paper.

\begin{figure}[t]
	\centering
	\begin{align*}
	\tilde{G}_{0}^{(2)}(P)&=
	\parbox[c]{7.5em}{
	\includegraphics[width=7.5em]{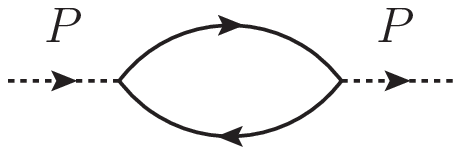}}
	\\
	\tilde{G}_{0}^{(3)}(P_1,P_2)&=
	\parbox[c]{7em}{
	\includegraphics[width=7em]{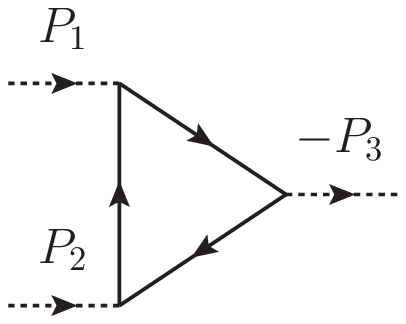}}
	+
	\parbox[c]{7em}{
	\includegraphics[width=7em]{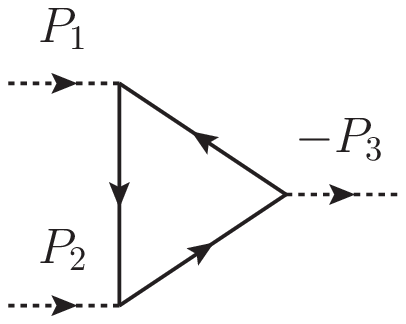}}
	\\
	\tilde{G}_{0}^{(4)}(P_1,P_2,P_3)&=
	\parbox[c]{7em}{
	\includegraphics[width=7em]{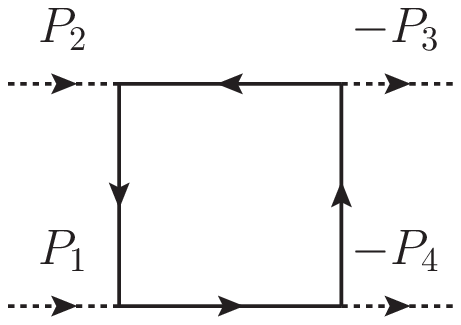}}
	+
	\parbox[c]{7em}{
	\includegraphics[width=7em]{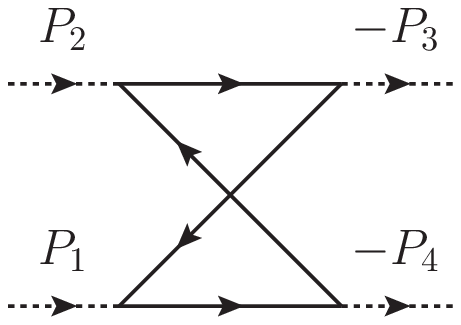}}
	\\
	&+
	\parbox[c]{7em}{
	\includegraphics[width=7em]{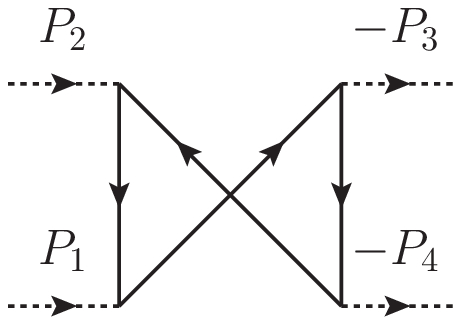}}
	+
	\parbox[c]{7em}{
	\includegraphics[width=7em]{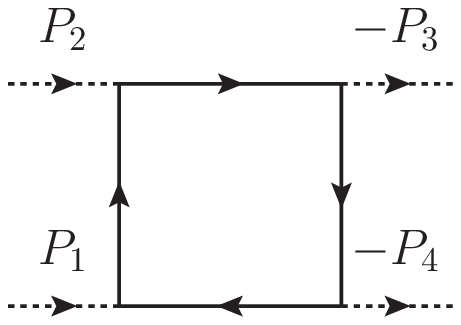}}
	\\
	&+
	\parbox[c]{7em}{
	\includegraphics[width=7em]{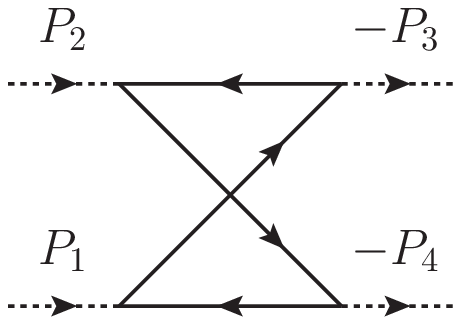}}
	+
	\parbox[c]{7em}{
	\includegraphics[width=7em]{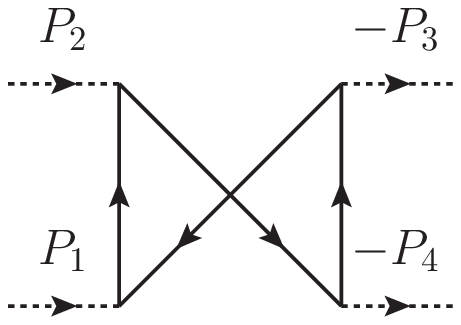}}
	\end{align*}
	\caption{The diagrammatic representations
	of $\tilde{G}_{0}^{(2,3,4)}$.
	The solid line is the free fermion propagator
	$\tilde{G}_{\rm F, 0}^{(2)}(P)$.
	In the diagrams for $\tilde{G}_{0}^{(3)}$ and $\tilde{G}_{0}^{(4)}$,
	$P_3=-P_1-P_2$ and $P_4=-P_1-P_2-P_3$, respectively.
	\label{fig:freediagram}}
\end{figure}
We specify the initial conditions
$\overline{E}_{\rm gs, \lambda=0}$,
$\rho_{\rm gs,\lambda=0}$,
and $\tilde{G}^{(2,3,4)}_{\lambda=0}$.
We denote $\rho_{\rm gs,0}$ as $n$, which is the ground state density
during the flow, and in particular at $\lambda=1$, because $\rho_{\rm gs,\lambda}(\chi)=\rho_{\rm gs,0}$.
The fermi momentum and fermi energy
are defined as $p_{\rm F}=\pi n$
and $E_{\rm F}=p_{\rm F}^2/2$, respectively.
$\overline{E}_{\rm gs, \lambda=0}$
is the ground state energy per particle
of a one-dimensional free Fermi gas:\,$\overline{E}_{\rm gs,0}=E_{\rm F}/3$.
$\tilde{G}^{(2,3,4)}_{\lambda=0}$ are the density correlation functions for free fermions:
\begin{align*}
	\tilde{G}^{(2)}_{0}(P)
	=&
	-
	\int_{P'}
	\tilde{G}_{\rm F, 0}^{(2)}(P')
	\tilde{G}_{\rm F, 0}^{(2)}(P+P'),
	\\
	\tilde{G}^{(3)}_{0}(P_1,P_2)
	=&
	-
	\sum_{\sigma\in S_{2}}
	\int_{P'}
	\tilde{G}_{\rm F, 0}^{(2)}(P')
	\tilde{G}_{\rm F, 0}^{(2)}(P_{\sigma(1)}+P')
	\\
	&\times
	\tilde{G}_{\rm F, 0}^{(2)}(P_{\sigma(1)}+P_{\sigma(2)}+P'),
	\\
	\tilde{G}^{(4)}_{0}(P_1,P_2,P_3)
	=&
	-
	\sum_{\sigma\in S_{3}}
	\int_{P'}
	\tilde{G}_{\rm F, 0}^{(2)}(P')
	\tilde{G}_{\rm F, 0}^{(2)}(P_{\sigma(1)}+P')
	\notag
	\\
	&
	\times
	\tilde{G}_{\rm F, 0}^{(2)}(P_{\sigma(1)}+P_{\sigma(2)}+P')
	\notag
	\\
	&\times
	\tilde{G}_{\rm F, 0}^{(2)}(P_{\sigma(1)}+P_{\sigma(2)}+P_{\sigma(3)}+P').
\end{align*}
Here $S_{2}$ and $S_{3}$ are the symmetric groups of order two and three, 
respectively,
and $\tilde{G}_{\rm F, 0}^{(2)}(P)$ is the two-point propagator
of free fermions: $\tilde{G}_{\rm F, 0}^{(2)}(P)={1}/(i\omega-\xi(p))$,
where $\xi(p):=p^2/2-E_{\rm F}$.
The diagrammatic representations of 
these correlation functions are shown in
Fig.~\ref{fig:freediagram}.
The explicit forms of $\tilde{G}^{(2)}_0(P)$
and the second and third terms of the right-hand side
of Eq.~\eqref{eq:gtfl} after the frequency integral
under the approximation Eq.~\eqref{eq:appr} becomes as follows:
\begin{widetext}
\begin{align}
	&\tilde{G}^{(2)}_{0}(P)
	=
	\int_{p'}
	\frac{2\theta(\xi(p+p'))\theta(-\xi(p'))
	\left(\xi(p+p')-\xi(p')\right)}{\omega^2+[\xi(p+p')-\xi(p')]^2},
	\label{eq:g02e}
	\\
	&
	-\frac{c_\lambda}{2}
	\int_{P'}
	\tilde{U}(p')
	\tilde{G}^{(4)}_{0}(P',-P',P)
	+
	c_\lambda
	\int_{P'}
	\frac{
	\tilde{U}(p')
	\tilde{G}_{0}^{(3)}(P',-P')
	\tilde{G}_{0}^{(3)}(P,-P)
	}{2\tilde{G}_{0}^{(2)}(0)}
	\notag
	\\
	&=
	2c_\lambda
	\int_{p',p''}
	\tilde{U}(p')
	\theta(-\xi(p' + p''))
	(\theta(-\xi(p''+p))-\theta(-\xi(p'')))
	\left[
	\frac{(\xi(p''+p)-\xi(p''))^2-\omega^2}{(\omega^2+(\xi(p''+p)-\xi(p''))^2)^2}
	\right.
	\notag
	\\
	&\quad
	+
	\left.
	\frac{2\theta(\xi(p''+p+p'))}{\xi(p''+p'+p)-\xi(p'' +p)-\xi(p''+p')+\xi(p'')}
	\frac{\xi(p''+p+p')-\xi(p''+p')}{\omega^2+(\xi(p''+p+p')-\xi(p''+p'))^2}
	\right].
	\label{eq:g4ue}
\end{align}
\end{widetext}
The flow equations \eqref{eq:eflow} and \eqref{eq:gtfl}
under the approximation Eq.~\eqref{eq:appr} with the expression
Eq.~\eqref{eq:g4ue} are found to be the same as those obtained from the continuum limit
of the system with finite number of particles in a finite box presented in \cite{kem17a}.
Finally we comment on the derivation of Eq.~\eqref{eq:g4ue}.
Both terms in the left-hand side of Eq.~\eqref{eq:g4ue} have
delta functions in the momentum integrals, 
which comes from the derivatives of the distribution functions
of fermions such as $\theta'(-\xi(p'))=-\delta(\xi(p'))$.
These contributions from the delta functions, 
however, cancel each other out and do not appear in the final expression
in Eq.~\eqref{eq:g4ue}.

\section{Demonstration in one-dimensional spinless nuclear matter\label{sec:demo}}
In this section, we demonstrate the calculation of 
the ground state energy as a function of the density,
i.e. the equation of state (EOS)
in a $(1+1)$-dimensional spinless continuum 
nuclear matter~\cite{ale89}.

\subsection{One-dimensional spinless nuclear matter}
We consider a $(1+1)$-dimensional continuum matter 
composed of spinless fermions with
the following two-body interaction~\cite{ale89}:
\begin{align*}
	U(r)
	=\frac{g}{\sqrt{\pi}}
	\left(
	\frac{1}{\sigma_1}{\rm e}^{-\frac{r^2}{\sigma_1^2}}
	-
	\frac{1}{\sigma_2}{\rm e}^{-\frac{r^2}{\sigma_2^2}}
	\right)
\end{align*}
where $\sigma_1 >0$, $\sigma_2 >0$ and $g >0$.
In this model, the short-range repulsive force and
long-range attractive force between nucleons
are represented with the superposition of the two Gaussians.
Following Ref.~\cite{ale89}, 
we choose $g=12$, $\sigma_1=0.2$ and $\sigma_2=0.8$
in units such that the mass of a nucleon is 1.	
These parameters are determined under the assumption that
some dimensionless quantities
obtained in the $(3+1)$-dimensional system empirically
are reproduced also in $(1+1)$ dimension.

In this model, the ground state energy 
of the continuum nuclear matter
is derived from the extrapolation of 
the ground state energies of the systems with 2 to 12 nucleons
calculated by the Monte Carlo (MC) simulation
at a given density which seems to be close
to the saturation density~\cite{ale89}.
Although there may be some discrepancy
between the energy derived from the MC simulation
and the saturation energy
due to the choice of the density,
we use the energy derived from the MC simulation
as a reference of the saturation energy
to benchmark our result.

\subsection{Numerical procedure}
We mention some details of our numerical analysis to solve the flow equations.

Equations~\eqref{eq:g4ue} has 
a seemingly singular point at $p'=0$
in the term proportional to $(\xi(p''+p'+p)-\xi(p''+p)-\xi(p''+p')+\xi(p''))^{-1}\sim p'^{-1}$ in the integrand.
This singularity, however, vanishes because $\tilde{U}(p'=0)=0$ for our interaction.
To avoid the division-by-zero operation, we rewrite the integrand as
a manifestly regular form for the numerical calculation by use of the Maclaurin expansion of $\tilde{U}(p')$.

In order to calculate $\tilde{G}^{(2)}_{\lambda}(\omega,p)$
on the $(\omega,p)$-plane, we discretize $\omega$ and $p$.
We change $\omega$ to $\bar{\omega}=(2/\pi)\arctan(\omega/s)$,
where $s$ is an arbitral number,
and discretize $\bar{\omega}$ in the domain $[-1,1]$.
We set a cutoff $\Lambda=\max (10p_{\rm F},1)$ 
for momentum $p$.
We have checked that the result hardly depends on $\Lambda$
even if it is set to larger values.

\subsection{Ground state properties}

\begin{figure}[!t]
	\centering
	\includegraphics[width=\columnwidth]{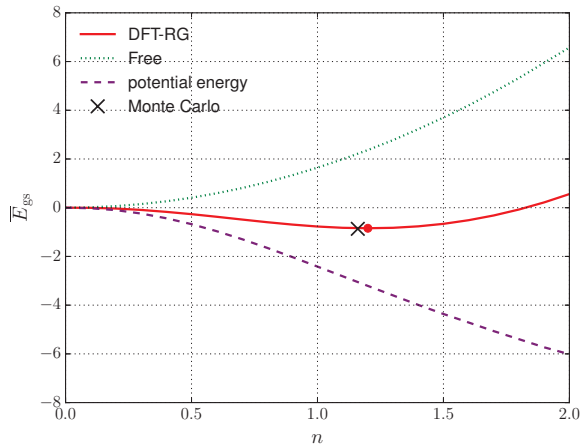}
	\caption{Energy per particle $\overline{E}_{\rm gs}$
	as a function of density $n$ in the case of
	DFT-RG (solid red line).
	The energy of free fermion (green dotted line),
	the contribution from the inter-nucleon
	potential (purple dashed line) and
	the result of the extrapolated energy given by
	a MC simulation~\cite{ale89} 
	at a density close but not equal 
	to saturation density (black cross)
	are also shown.
	The red point is the saturation point derived
	from the DFT-RG calculation.
	The units for $\overline{E}_{\rm gs}$ and $n$ are such that the mass of a nucleon is $1$.
	\label{fig:energy}}
\end{figure}
\begin{table*}[!htb]
	\begin{center}
		\caption{The saturation energies $\overline{E}_{\rm s}$
		derived from the first-order perturbation,
		DFT-RG with $c_\lambda=1$, DFT-RG,  and 
		the MC simulation~\cite{ale89}.
		Their relative errors compared to the result
		of the Monte Carlo simulation $\overline{E}_{\rm s,MC}$
		and the saturation densities $\rho_{\rm s}$
		are also shown.
		The units for $\overline{E}_{\rm s}$ and $\rho_{\rm s}$ are such that the mass of a nucleon is $1$.
		\label{tab:saturation}
		}
		\begin{tabular}{ccccc}
		  \hline
		  & 1st order & DFT-RG ($c_\lambda=1$) & DFT-RG & Monte Carlo~\cite{ale89} \\
		  \hline
		  $\left|\overline{E}_{\rm s}\right|$ 
		  & 0.780 & 0.827 & 0.844 & 0.867 \\
		  $\left|\left(\overline{E}_{\rm s}-\overline{E}_{\rm s,MC}\right)
		  /\overline{E}_{\rm s,MC}\right|$ (\%)
		  & 10.0 & 4.6 & 2.7 & - \\
		  $\rho_{\rm s}$ & 1.19 & 1.21 & 1.20 & -\\
		  \hline
		\end{tabular}			
	\end{center}
\end{table*}

We show our result of the EOS
together with the energy of the free Fermi gas 
$\overline{E}_{\rm gs,0}$
and 
the contribution from the inter-nucleon potential
$\overline{E}_{\rm gs,1}-\overline{E}_{\rm gs,0}$
in Fig.~\ref{fig:energy}.
The free case shows an increase with respect to $n$, 
which reflects that 
the average kinetic energy increases 
because the fermi sphere becomes larger as $n$ increases.
On the other hand, 
the contribution from the inter-particle potential reduces
the energy because it has a long-range attractive part.
The competition between these contributions results in 
the emergence of a minimum point at a finite $n$, 
i.e. the saturation point.

\begin{figure}[!t]
	\centering
	\includegraphics[width=\columnwidth]{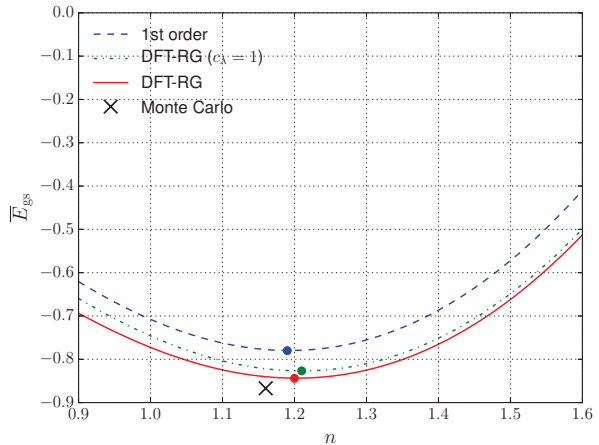}
	\caption{Energy per particle $\overline{E}_{\rm gs}$
	as a function of density $n$ near the saturation point.
	The result of
	DFT-RG (solid red line), 
	first-order perturbation (blue dashed line),
	DFT-RG with $c_\lambda=1$ (green dotted--dashed line), 
	and a MC simulation~\cite{ale89} (black cross)
	are shown.
	The density $n = 1.16$ used in the MC simulation
	is an assumed one which is close but not equal 
	to saturation density.
	The units for $\overline{E}_{\rm gs}$ and $n$ are such that the mass of a nucleon is $1$.
	\label{fig:energy_detail}}
\end{figure}

To discuss the quantitative aspects of our result,
we show our result of the EOS near the
saturation point in Fig.~\ref{fig:energy_detail}.
For comparison, the results with
the first-order perturbation 
and with ignoring the running of $c_\lambda$,
i.e. $c_\lambda=1$, are also shown.
In our formalism, the first-order perturbation
is reproduced when $\tilde{G}^{(2)}_\lambda(P)$
in Eq.~\eqref{eq:eflow} is replaced with
$\tilde{G}^{(2)}_{0}(P)$.
In Fig.~\ref{fig:energy_detail},
one finds that the saturation energy derived from DFT-RG
is closer to the result of the MC simulation
than any other methods.
We again note that the density used in
the MC result shown in Fig.~\ref{fig:energy_detail} 
is not a calculated but given density 
which is assumed to be close but not equal to the saturation density.
Therefore, the density used in the MC simulation
should not be used as a reference of the benchmark
of the saturation density.

Table~\ref{tab:saturation} shows the numerical
results of the saturation energy $\overline{E}_{\rm s}$
and density $\rho_{\rm s}$ derived from each method.
As shown in this table, 
the deviation of the saturation energy 
between DFT-RG and the MC simulation
is only 2.7\%.
Compared to the result of the first-order perturbation,
the accuracy of $\overline{E}_{\rm s}$
is largely improved by use of the DFT-RG scheme.
Moreover, the introduction of the running of 
the factor $c_\lambda$ also contributes to 
the improvement of the accuracy.
Although there is no reference for the saturation
density $\rho_{\rm s}$,
it seems to converge to a value near $1.20$.

\section{Conclusion\label{sec:conc}}
In this paper,
we have presented the functional renormalization 
group (FRG)-aided density-functional theory (DFT) calculation 
of the equation of state of an infinite nuclear matter
in (1+1)-dimensions composed of spinless nucleons.
We have shown a novel formalism to treat infinite matters
in which the flow of the chemical potential is taken into account
to control the particle number during the flow.
The resultant saturation energy
coincides with that obtained from 
the Monte-Carlo simulation within a few percent. 
Thus one sees that DFT-RG scheme works well 
for the infinite homogeneous nuclear 
system around the saturation point 
in comparison with the case of finite nuclear system \cite{kem17a}
where the same approximation 
for the flow equations was employed:
A truncation including 
up to the second-order vertex expansion was used
and an approximation was made for the three- and four-point correlation functions so that the
Pauli blocking effect is taken into account.

Our result together with its numerical feasibility
clearly demonstrate that the DFT-RG scheme is a promising nonperturbative method
to analyze quantum many-body systems, at least,  composed of infinite number of particles.
One of the next steps is applying our method 
to higher-dimensional systems which should be easy in
 the present grand canonical formalism.
The extension to a finite-temperature case can be done as well.
The extension to particles with internal 
degrees of freedom \cite{kem17b}
is also an important future direction.

Another interesting extension of the present work
is calculation of dynamical quantities.
One of the basic quantities for analyzing the dynamical properties
is the spectral functions.
Recently, the FRG has been developed so that  
 the spectral functions can be calculated which describe
 the dynamical properties of the system 
\cite{kam14,tri14a,tri14b,yok16,yok17}.
It is intriguing to extend the DFT-RG scheme so that 
the spectral function can be calculated.
We have already calculated the spectral function of the density--density correlation
function in the DFT-RG scheme for the one-dimensional model used in this work.
We will report the analysis of the spectral function
in a separate paper, where some relevance to the Tomonaga-Luttinger liquid 
with a non-linear dispersion relation \cite{ima12} will be discussed.
In conclusion, the present work demonstrates that 
the FRG-aided DFT scheme can be as powerful as any other 
methods in quantum many-body theory. 

\section*{Acknowledgments}
We thank Jean-Paul Blaizot for his interest in and
critical and valuable comments on the present work.
We also acknowledge Christof Wetterich, Jan M. Pawlowski and Jochen Wambach
for their interest in and fruitful comments on the present work.
T.~Y. was supported by the Grants-in-Aid for JSPS fellows
(Grant No. 16J08574).
K.~Y. was supported by the JSPS KAKENHI (Grant No. 16K17687). 
T.~K. was supported by the JSPS KAKENHI Grants (Nos. 16K05350 and 15H03663)
and by the Yukawa International Program for Quark-Hadron Sciences (YIPQS).

\end{document}